\documentclass[12pt]{iopart}

\def\bete{{\beta\epsilon}}
\def\bacol{\setlength{\arraycolsep}{0pt}} 
\def\ds{\displaystyle}
\def\bec{\begin{center}} 
\def\enc{\end{center}} 

\def\ba{\begin{array}}
 
\def\ea{\end{array}} 
\def\btab{\begin{table}}
\def\btabu{\begin{tabular}} 
\def\etab{\end{table}}
\def\etabu{\end{tabular}} 
\def\bit{\begin{itemize}}
\def\eit{\end{itemize}} 
\def\bef{\begin{figure}[htb]}
\def\befh{\begin{figure}[!h!]}  
\def\enf{\end{figure}}
 
\def\la{\langle} 
\def\ra{\rangle}
\def\pd{\partial}

\def\bzero{\mbox{\boldmath $0$}}
\def\hK{{\hat K}}

\def\d{\delta} 
\def\e{\epsilon}

\def\L{\Lambda}

\def\O{\Omega}

\def\s{\sigma}

\def\cD{{\cal D}}

\def\cL{{\cal L}}

\def\half{{\textstyle{1 \over 2}}}

\def\b1{{\bf 1}}

\def\bx{{\bf x}} 
\def\bk{{\bf k}}
\def\bfa{{\mbox{\boldmath $a$}}} 
\def\bc{{\mbox{\boldmath $c$}}}

\def\bF{\mbox{\boldmath $F$}}
\def\bG{\mbox{\boldmath $G$}} 
\def\bD{\mbox{\boldmath $D$}}
 
\def\bX{\mbox{\boldmath $X$}}
\def\bB{\mbox{\boldmath $B$}} 
\def\bE{\mbox{\boldmath $E$}}

\def\cos{\;\hbox{cos}} 
\def\sin{\;\hbox{sin}}

\def\sign{\hbox{sign}} 

\def\nn{\nonumber}

\def\bs{{\bf s}}

\def\bs{{\bf s}}

\def\tphi{{\tilde{\phi}}} 
\def\txi{{\tilde{\xi}}}

 \usepackage{epsfig,latexsym,subfigure}

\newcommand \bew {\begin{widetext}} 
\newcommand \enw {\end{widetext}}

\begin{document}

\title[Thermal Casimir effect in layered systems]
{The field theory of symmetrical layered
electrolytic systems and the thermal Casimir effect}

\author{DS Dean\dag\  and RR Horgan\ddag}

\address{\dag\  Laboratoire de Physique Th\'eorique, CNRS UMR5152,
  IRSAMC, Universit\'e Paul Sabatier, 118 route de Narbonne, 31062
  Toulouse Cedex 04, France}

\address{\ddag\ DAMTP, CMS, University of Cambridge, Cambridge, CB3
  0WA, UK}

\begin{abstract}
We present a general extension of a field-theoretic approach developed
in earlier papers to the calculation of the free energy of
symmetrically layered electrolytic systems which is based on the
Sine-Gordon field theory for the Coulomb gas. The method is to
construct the partition function in terms of the Feynman evolution
kernel in the Euclidean time variable associated with the coordinate
normal to the surfaces defining the layered structure. The theory is
applicable to cylindrical systems and its development is motivated by
the possibility that a static van der Waals or thermal Casimir force
could provide an attractive force stabilising a dielectric tube formed
from a lipid bilayer, an example of which are t-tubules occurring in
certain muscle cells.  In this context, we apply the theory to the
calculation of the thermal Casimir effect for a dielectric tube of
radius $R$ and thickness $\delta$ formed from such a membrane in
water. In a grand canonical approach we find that the leading
contribution to the Casimir energy behaves like $-k_BTL\kappa_C/R$
which gives rise to an attractive force which tends to contract the
tube radius. We find that $\kappa_C \sim 0.3$ for the case of typical
lipid membrane t-tubules. We conclude that except in the case of a
very soft membrane this force is insufficient to stabilise such tubes
against the bending stress which tend to increase the radius. We
briefly discuss the role of lipid membrane reservoir implicit in the
approach and whether its nature in biological systems may possibly
lead to a stabilising mechanism for such lipid tubes.
\end{abstract}  

\section{Introduction}
In an recent short communication we reported on a calculation which
investigated the the possibility that a static van der Waals or
thermal Casimir force could provide an attractive force across a tube
formed from a lipid bilayer, so leading to its stabilisation.  In this
paper we give the details of the general theory of symmetrically
layered electrolytic systems which underlies that calculation, and
explain the details of the calculation applying the theory to
cylindrical geometry and to a model for the lipid bilayer tube. Whilst
the motivation for developing the theory presented below is the
analysis of the the Casimir force in the context of a dielectric tube
immersed in water, the theory is applicable to any sufficiently
symmetrical system consisting of layer containing electrolyte. The
Coulomb properties of such systems are described by a Sine-Gordon
field theory and a full analysis in the case of flat layers has been
done with the approach which is generalised in this paper. In
particular, it allows for the perturbation series for the thermal
Casimir force to be developed in terms of the dimensionless coupling
$g=l_B/l_D$ where $l_B$ and $l_D$ are the Bjerrum and Debye lengths,
respectively.

The behaviour of systems composed of layer of varying dielectric
constants was first studied by Lifshitz and coworkers \cite{lif}
and has been subsequently revisited by a number of authors
\cite{mani,netz,deho}. The formalism
developed is an elegant way of taking into account van der Waals
forces in a continuum theory. Two types of van der Waals forces are
accounted for in theses theories, firstly zero frequency van der Waals
forces whose nature is purely classical and secondly the frequency dependent
ones due to temporal dipole fluctuations. In terms of thermal field
theory the former correspond to the zero frequency Matsubara frequency
and the latter to the non zero frequencies. In order to 
calculate these latter terms we require
information about the frequency dependence of the dielectric
constants, where as the former only requires the static dielectric
permittivity.  The quantum Casimir effect corresponding to the
modification of the ground state energy of the electromagnetic field
has been intensively studied in the case of idealised boundary
conditions in a variety of geometries including spheres and cylinders
\cite{milton}. The thermal Casimir effect investigated here has a
similar mathematical structure though the corresponding effective
spatial dimension is one less in the calculations. The temperature
dependence of the full Casimir in a simplified model of a solid
dielectric cylinder (and sphere) has been recently examined using a
heat kernel coefficient expansion \cite{bonepi}.  In our analysis of
the diffuse limits we make use of summation theorems for Bessel
functions which were introduced for the study of the Casimir energy
for cylinders with light-velocity conserving boundary conditions
\cite{klro}. In this paper we will calculate only the zero frequency
contribution, also known as the thermal Casimir effect. The thermal
Casimir effect may also be calculated in the presence of electrolyte and 
the technique we develop here for electrolytic systems within the
Debye H\"uckel approximation is valid in the domain of weak electrolyte
concentrations. There is an extensive literature on the thermal
Casimir effect for systems of layered geometries, both without added
electrolyte and within the Debye H\"uckel approximation \cite{mani,netz,deho}. 
Recently the calculation of the thermal Casimir force for layered films at the
first order of perturbation theory about the Debye H\"uckel theory was
carried out \cite{dehotloop}, suggesting the possibility of strong non
perturbative effects.

In section \ref{ttube} we discuss the model for the lipid bilayer tube
and review the outcome of the calculation applied to this model; in
section \ref{theory} we present the general theory for calculating the
free energy of a general symmetrically layered electrolytic system; in
section \ref{cylinder_casimir} we apply the general theory to
dielectric layers with cylindrical geometry; in section
\ref{casimir_eval} we present the calculation of the Casimir force for
the particular case of a dielectric tube of thickness $\delta$ and
radius $R$ immersed in water; in section \ref{numbers} we evaluate the
Casimir force for physically reasonable values of $\delta,R$ and in
section \ref{conclusion} we present some conclusions.

\section{\label{ttube}The lipid tubule}
The behaviour of lipid bilayers is of crucial importance in
biophysics. Lipid bilayers in water exhibit a huge variety of
geometries and structures and in the context of cell biology even more
varied structures are exhibited. In order to understand where
biological mechanisms such as molecular motors and cytoskeletal
structures are determinant in the stability of biological structures,
one must first understand the role of the basic physical interactions
in systems that contain only lipid bilayers, {\em i.e.} model membrane
systems.  There has been much study of lipid bilayer shape and
elasticity using standard continuum mechanics \cite{helfa,boal}. This basic
approach is also complemented by more microscopic studies based on
lipid structure and lipid-lipid interaction models, this approach is
of course ultimately necessary to fully understand the physics of
bilayers.  The bilayer is composed of two layers of lipid each layer
having the hydrophilic lipid head at the surface where it is in
contact with water, the interior is composed of the lipid's
hydrocarbon tails. This layer geometry is stable due to the
hydrophobic nature of the hydrocarbon tails. Given this
non-homogeneous structure one can immediately see that a simple continuum
elastic sheet type model may have difficulty in predicting the
mechanical properties of bilayers.

In certain muscle cells, structures known as t-tubules are
found. These are basically cylindrical tubes whose surface is composed
of a lipid bilayer. Similar structures may also be mechanically drawn
off bilayer vesicles. The stability of these tubular structures
requires an explanation. The basic continuum theory\cite{helfa,boal} 
predicts that the
free energy of a tube of length $L$ and radius $R$ is
\begin{equation}
F_B(L,R) = \frac{k_BTL\kappa_B}{R}\;,
\end{equation}
where the above expression is strictly speaking the excess free energy
with respect to a flat membrane of the same area $A=2\pi RL$ and the
subscript $B$ refers to mechanical bending. Various experimental and
theoretical estimates for $\kappa_B$ can be found in the literature
\cite{boal,wurg}
and they lie between $3$ and $30$. Note that our definition of $\kappa_B$
differs from that used traditionally in the literature $\kappa'_B$ by a 
factor of $\pi$, $\kappa_B = \pi \kappa'_B$.
The values of $\kappa_B$ depend of course on the composition of the 
bilayer and on the experimental protocol used to measure it. 
One crucial element in both
theoretical and experimental determinations of $\kappa_B$ is whether
the tube is attached to a reservoir of lipid or not, {\em i.e.}
whether the statistical ensemble is grand canonical or grand
canonical. Clearly if there is no reservoir then any increase in the
surface area of the tube will lead to a less dense lipid surface
concentration, in this case water may be able to become in contact
with the internal layer composed of the hydrophobic heads and thus a
significant increase in free energy. If upon, changing the area of the
tube, lipids can flow into the tube to maintain the local optimal
packing then the free energy cost will be substantially different.
This bending free energy is positive and hence the preferred
thermodynamic state is the flat one. 
Mechanical models for membranes vary in their predictions for the 
dependence of $\kappa_B$ on the membrane thickness $\delta$. The models
most compatible with the available experimental data predict that
\begin{equation}
\kappa_B = K_A \delta^2/\alpha
\end{equation}
where $\alpha$ depends on the precise model and is generically  $O(10)$
and $K_A$ is the area compression modulus \cite{boal,rac}. Experimental
fits of $\kappa_B$ with respect to the membrane thickness are compatible 
with
\begin{equation}
\kappa_B = K_A (\delta-\delta_0)^2/\alpha
\end{equation}
where $\delta_0\approx 1$nm is an offset necessary to fit the data.  
We note that when lipid tubes are
drawn from a vesicle the mechanically applied tension can of course
overcome this free energy barrier. A natural question motivated by the
fact we see these structures in cells is whether there are any other
mechanisms that could lead to their formation and explain their
stability. A possible explanation is that
electrostatic effects involving surface charges and ions (salt) in the
surrounding medium could play a role\cite{degen,wsalt,chjasi} . 
Certain experiments \cite{wsalt} however
revealed a relative insensitivity of some system to the concentration of
salt. There are however other systems where the salt concentration
does appear important in determining the stability of the
tubules \cite{ssalt}, in these systems the lipid head groups are 
highly charged. Another explanation has been put forward in terms of the
geometry of the lipid, notably the tail having a structure such that
there is a preferred orientation of the tails next to each other,
giving rise to a chirality which allows the stabilisation of the
tubes \cite{hepr,sesc,cen,fope,dejupr}.. This explanation would however 
depend on a more or less mono-disperse lipid bilayer in order to permit 
this liquid crystal like
phase. Cell membranes are composed of a wide variety of lipid types
and addition have proteins present and so it is possible that another
mechanism is responsible for the stability of these structures

We adopt a continuum model where the lipid bilayer is modelled as a
layer of thickness $\delta \approx 5-10\mbox{(nm)}$ and of dielectric
constant $\epsilon'\approx 2\epsilon_0$.  The surrounding water is
also treated as a dielectric continuum of dielectric constant
$\epsilon \approx 80 \epsilon_0$. We shall also adopt a model where
the lipid tube is fixed at each end to a flat lipid reservoir and thus
work in the grand canonical ensemble; this is shown schematically in
Fig. \ref{tube1}.

In this paper we find that the thermal Casimir effect gives a
contribution to the excess free energy above the flat plane of
\begin{equation}
F_C(L,R) = -\frac{k_BTL\kappa_C}{R} \label{Fandkc}
\end{equation}
with
\begin{equation}
\kappa_C~=~\frac{\Delta^2}{64}\left[3\log\left(\frac{\pi\delta}{a}\right)
+ 6\log 2 + 3\gamma_E - 4\right] + \Delta^4B(\Delta)\;, \label{eqkc}
\end{equation}
where $a$ is a microscopic cut-off corresponding to the molecular/
lipid size below which the continuum picture of the dielectric medium
breaks down. We note that the sign of $F_C$ has exactly the same
functional form as the bending free energy $F_B$ but is of opposite sign, 
meaning that this force tends to collapse the tube and thus helps to 
stabilise the
system against the bending energy. We shall show later that with
reasonable physical parameters $\kappa_C \approx 0.5-1.0 $. Thus the Casimir
attraction is not able to overcome the repulsion due bending as it is
predicted by current theories and data. However this result is
important for several reasons:
\begin{itemize}
\item We show that the Casimir attraction tends to stabilise the tube
structure.
\item The presence of the microscopic cut off in $\kappa_C$ shows that
the physics is ultimately dominated by the short scale or ultra-violet
physics. This means that weak electrolyte concentrations will have
little effect on the system as seen in experiments given that there
are no strong surface charges.
\item We see that $F_C$ and $F_B$ have the same functional
form at large $R$ and that the behaviour of $\kappa_C$ is regulated by 
the microscopic physics. This means that our calculation can be interpreted
as a renormalisation of $\kappa_B$ due to the thermal Casimir 
effect. 

\item Further attractive,or tube stabilising, interactions may be
generated by the high frequency Matsubara modes.

\end{itemize}
\section{\label{theory} The Schr\"odinger kernel for separable systems}
The mathematical tool that allows us to derive the free energy for
electrolytic systems with symmetrical layered films is the functional
Schr\"odinger kernel which evolves the Sine-Gordon scalar field from
some in initial surface to a final surface.  Our method is applicable
when the Laplacian is separable in the natural coordinates describing
the surfaces bounding the layers of the system. In the surface between
the bounding layers the electrostatic and chemical properties of the system
are uniform {\em i.e.} the dielectric constants and electrolyte concentrations
are constant.
It is in this sense
that we describe such a system as symmetrical. In this case, in
$D$-dimensions, the coordinates can be denoted by $(\bx,\s)$ where
$\bx$ is a list of $(D-1)$ coordinates for surfaces $\s =
\mbox{constant}$.  The $i$-th surface of an $N$-layer system is
described by $\s=\s_i$, where the $\s_i$ are constants with $\s_{i+1}
> \s_i,~0\le i \le N$ with $\s_0$ and $\s_N$ being respectively the
minimum and maximum values in the range of $\s$. The local electrochemical
properties of the system thus depend solely on the coordinate $\s$. 
Our example in this
paper will be that of coaxial cylinders in $D=3$, where $\bx =
(\theta,\phi)$ and $\s=r$, the radius. However, the theory is more
general than for cylindrical or spherical coordinates, and so we lay
the theory out below in a general notation but refer to the
cylindrical case for clarity where appropriate.  The dynamics of the
field $\phi(\bx,\s)$ are defined by its evolution in the Euclidean
time coordinate $t,~-\infty < t < \infty$ which is given in terms of
$\s$. The volume measure is $dv = J(\s)d\s d\bx$ and the Euclidean
time $t(\s)$ is defined by 
\begin{equation}
t(\s_2)-t(\s_1)~=~\int_{\s_1}^{\s_2}\;\frac{d\s}{J(\s)}\;.\label{etime}
\end{equation} 
For example, in the cylindrical geometry $\s =r,~t = \log \s$ and
in the planar case $t=\s=z$.

To derive the general form for the kernel it is convenient to express
the contribution from one layer of the system to the total free energy
in dimensionless variables. In a previous paper \cite{dehotloop} we
derived an expression for the grand partition function of a layered
system in a dimensionless form, and in the present context the
effective action is the Sine-Gordon field theory which defines the
kernel and is written as 
\begin{equation} {\cal S} = -{1\over 8\pi}\int_{V_{12}}
dv\:\left(\nabla\phi\right)^2 + {Z(g)\over 4\pi g} \int_{V_{12}} dv\:
\cos(\sqrt{g} \phi), \label{S_L} 
\end{equation} 
where the region defining the
layer is bounded by two neighbouring surfaces $S_1$ and $S_2$ defined,
respectively, by $\s=\s_1,~\s=\s_2,~\s_1<\s_2$, and has volume
$V_{12}$. All lengths are measured in terms of the Debye length, $l_D
= 1/m$, where $m = \sqrt{2 \rho e^2 \beta/\epsilon}$ is the Debye
mass, $\rho$ is the ion density of the bulk reservoir to which the
electrolyte solution within the layer is connected, and $\epsilon$ is
the dielectric constant in the layer. The other fundamental length in
the theory is the Bjerrum length, $l_B = e^2 \beta/4\pi \epsilon$, and
the dimensionless coupling constant is given by $g = l_B/l_D$.  The
dimensionful field is given in terms of $\phi$ by the rescaling 
\begin{equation}
\phi~\longrightarrow~\frac{e\beta}{\sqrt{g}}\phi~. \label{phi_rscale}
\end{equation} 
The renormalisation constant $Z(g)$ is associated with the ion
chemical potential $\mu$ conjugate to $\rho$, and removes the
divergences due to the unphysical charge self-interactions. In
Eq. (\ref{S_L}), $\mu$ has been substituted by the reservoir density
$\rho$ using the relation 
\begin{equation}
\mu~=~Z(g)\rho\;,~~~~~~Z(g)~=~\frac{1}{\langle
\cos(\sqrt{g}\phi)\rangle_B} \label{eqzg}\;, \end{equation} 
where the
above subscript $B$ indicates that expectation value is for an 
infinite bulk system.

The total partition function is constructed as a convolution of the
kernels of the layers in sequence, and to carry this out the
dimensionful description must be restored. For multiple layers the
action is a sum of similar terms each associated with a layer of the
system bounded by an inner and an outer surface.  In particular, the
innermost and outermost surfaces are at $\s_0$ and $\s_N$
corresponding to $t=-\infty$ and $t=\infty$, respectively. It was
shown in \cite{deho,dehocon} that for planar interfaces the
Schr\"odinger kernels which are bounded by one or the other of these
surfaces are given in terms of the ground-state wave-function of the
appropriate free Hamiltonian, and that this is sufficient to ensure
that the overall charge neutrality constraint is respected. In the
more general case, where the interfaces are non-planar (cylindrical,
for example), the Hamiltonian depends explicitly on the Euclidean time
$t$ and so there is no interpretation in terms of stationary
eigenstates. However, in the limit $t \to \pm\infty$ the relevant
kernels are separable in the boundary fields, and this leads to the
same result.

The action $S$ in Eq. (\ref{S_L}) can be decomposed as 
\begin{equation} {\cal S} =
{Z(g)\over 4\pi g}V + {\cal S}^{(0)} + \Delta {\cal S}\;, \end{equation} 
where
$V$ is the volume of the layer and the first term is the ideal
contribution. The term ${\cal S}^{(0)}$ is the action for a free or
Gaussian field theory and is given by 
\begin{equation} {\cal S}^{(0)} = -{1\over
8\pi}\int_V dv\:\left[ \left(\nabla\phi\right)^2 +
\phi^2\right]\;. \label{S0} 
\end{equation} 
The interacting part of ${\cal S}$ is
expressed as a perturbation 
\begin{equation} \Delta {\cal S} = {1\over 4\pi
g}\int_V dv\:\left[ Z(g)\left(\cos(\sqrt{g}\phi) -1\right) + {g
\phi^2\over 2}\right]\;, \label{DS} 
\end{equation} 
and the action ${\cal S}_B$
for the equivalent bulk system is given by 
\begin{equation} {\cal S}_B = -{1\over
8\pi}\int_B dv\: \left(\nabla\phi\right)^2 + {Z(g)\over 4\pi g} \int_B
\ dv\:\cos(\sqrt{g} \phi)\;, \label{S_B} 
\end{equation} 
which may be decomposed
in the same manner as for $S$.

The Schr\"odinger kernel for the layer is defined by 
\begin{equation}
\hK(\phi_2(\bx),\s_2;\phi_1(\bx),\s_1)~=~\int_{\phi_1}^{\phi_2}\;\cD\phi\;e^{S(\phi)}\;,\label{K_L}
\end{equation}
where $\phi_i(\bx) = \phi(\bx,\s_i),~i=1,2$, are the boundary
values of the field $\phi(\bx,\s)$ on the bounding surfaces $S_i$,
respectively.

In this section we concentrate on the calculation of
$\hK^{(0)}(\phi_2(\bx),\s_2;\phi_1(\bx),\s_1)$ defined by 
\begin{equation}
\hK^{(0)}(\phi_2(\bx),\s_2;\phi_1(\bx),\s_1)~=~\int_{\phi_1}^{\phi_2}\;\cD\phi\;e^{S^{(0)}(\phi)}\;.\label{K0_L}
\end{equation} 
The explicit evaluation of $\hK^{(0)}$ for the specified geometry
gives the Casimir-effect contribution from the layer to the free
energy $\O = -k_BT\log \Xi$, where $\Xi$ is the grand partition
function for the system, and forms the basis for a perturbative
expansion of $\O$ in terms of the interaction coupling strength
$g$. For an $N$-layer system the grand partition function for the free
theory, $\Xi^{(0)}$, is given by the convolution over layers as 
\begin{equation}
\Xi^{(0)}~=~\int\;\prod_{i=0}^{N}\cD\phi_i\;
\hK^{(0)}_i(\phi_{i+1}(\bx),\phi_i(\bx),\s_{i+1},\s_i)\;, \end{equation}
where
$t(\s_0)=-\infty, t(\s_N)=\infty$, and where the $\hK^{(0)}_i$ are
re-expressed in terms of the original, dimensionful, boundary fields
so that their values match correctly on the common interface
separating successive layers. The Casimir free energy is then given by 
\begin{equation}
F_C~=~\O^{(0)}-\O^{(0)}_B\;. 
\end{equation} 
Here  $\O^{(0)}_B$ is the equivalent
bulk contribution of an independent set of pure bulk systems having the 
same volume and properties as the layers composing the system. In this way the
generalised force corresponding to the position of any interface is a 
disjoining pressure.

We shall now show how to explicitly compute
$\hK^{(0)}(\phi_2,\s_2;\phi_1,\s_1)$ in its dimensionless form.  The
volume measure in Eq. (\ref{K0_L}) is $dv = J(\s)d\s d\bx$ where
$J(\s)$ is the Jacobian of the measure.  Since the functional integral
defining $\hK^{(0)}$ is Gaussian in form we explicitly find the
classical field $\phi_c$ which minimises the action by solving the
linear field equation 
\begin{equation} 
-(\nabla \cdot
J(\s)\nabla)\phi_c~+~J(\s)\phi_c~=~0\;, \label{feqn} 
\end{equation} 
with
boundary constraints 
\begin{equation}\phi_c(\bx,\s_1)~=~\phi_1(\bx),~~~~
\phi_c(\bx,\s_2)~=~\phi_2(\bx)\;. \label{bcnds} \end{equation} 
We assume that
the operator $\nabla \cdot J(\s)\nabla$ is separable which allows us
to write this field equation as 
\begin{equation}
-\frac{d}{d\s}J(\s)\frac{d}{d\s}\phi_c~-~J(\s)(\nabla^2_\bx+1)\phi_c~=~0\;,
\end{equation} 
where $\nabla^2_{\bx}$ is self-adjoint and may depend on $\s$ but
not on derivatives with respect to $\s$.  The orthonormal
eigenfunctions of $-\nabla^2_\bx$ are denoted $X(\bs,\bx)$ with
eigenvalue are $\lambda(\bs,\s)$: 
\begin{equation} -\nabla^2_\bx
X(\bs,\bx)~=~\lambda(\bs,\s) X(\bs,\bx)\;, \label{Xsx} \end{equation} 
where
$\bs$ is a set of $D-1$ quantum numbers. The classical field
$\phi_c(\bx,\s)$ is expanded on the complete set of functions $\{X\}$
as 
\begin{equation} 
\phi_c(\bx,\s)~=~\sum_\bs\;T(\bs,\s)X(\bs,\bx)\;, \label{PHIC}
\end{equation} 
where $ T(\bs,\s)$ satisfies the ordinary differential equation
\begin{equation}
\left[-\frac{d}{d\s}J(\s)\frac{d}{d\s}~+~J(\s)(\lambda(\bs,\s)+1)\right]
T(\bs,\s)~=~0\;. \label{Tst} 
\end{equation} 
We denote two 
solutions of this equation by $F_1(\bs,\s)$ and $F_2(\bs,\s)$, where
$F_1(\bs,\s)$ is finite as $t(\s) \to -\infty$ and $F_2(\bs,\s)$ is
finite as $t(\s) \to \infty$. 
In addition, these functions with
different quantum numbers $\bs$ are orthogonal with respect to  
the appropriate
measure. 
The Wronskian is given by the identity 
\begin{equation}
J(\s)[F_1(\bs,\s)F_2^\prime(\bs,\s) -
F_1^\prime(\bs,\s)F_2(\bs,\s)]~=~1\;. \label{wronskian} 
\end{equation} 
Then we
can write 
\begin{equation} T(\bs,\s)~=~a_1(\bs) F_1(\bs,\s)~+~a_2(\bs)
F_2(\bs,\s)\;. \label{T} 
\end{equation} 
The boundary fields $\phi_i$ on the
surfaces $S_i$ of the system can be expanded as 
\begin{equation}
\phi_i(\bx)~=~\sum_\bs\;c_i(\bs)X(\bs,\bx)\;,~~~~0 \le i \le N\;.
\label{PHII}
\end{equation} 
For the generic layer under discussion we consider the bounding
surfaces to be $S_1$ and $S_2$. Comparing with Eqs. (\ref{PHIC}) and
(\ref{T}), we find the relation between $\bc(\bs) =
(c_1(\bs),\;c_2(\bs))$ and $\bfa(\bs) = (a_1(\bs)\;,a_2(\bs))$ to be

\begin{equation} 
\fl \bc = \bfa\cdot\bF(\bs,\s_2,\s_1)\;,~~~~~~ \bF(\bs,\s_2,\s_1)~=~
\left( \ba{ccc} F_1(\bs,\s_1)&~~~~&F_1(\bs,\s_2)\\ &&\\
F_2(\bs,\s_1)&~~~~&F_2(\bs,\s_2) \ea \right)\;. \label{F} 
\end{equation} 
Now
using the classical field in Eq. (\ref{PHIC}) and the definition of
$S^{(0)}$ from Eq. (\ref{S0}) we find that the free classical action
is $S^{(0)}(\phi_c)$ is given by the boundary term 
\begin{equation}
S^{(0)}(\phi_c)~=~
-\frac{1}{8\pi}\int\;d\bx\;\left[J(\s)\phi_c(\bx,\s)\frac{d\phi_c(\bx,\s)}{d\s}\right]_{\s_1}^{\s_2}\;,\label{S0C}
\end{equation} 
where we have used integration by parts.

We use the expansion of $\phi_c(\bx,\s)$ in Eq. (\ref{PHIC}) in terms
of the coefficients $\bfa(\bs)$ and the expansion of
$\phi_i(\bx),~i=1,2$, in Eq. (\ref{PHII}) in terms of the coefficients
$\bc(\bs)$, and also use the fact that the functions of the basis set
$\{X(\bs,\bx)\}$ are orthonormal. We can then eliminate $\bfa(\bs)$ in
favour of $\bc(\bs)$, and find from Eq. (\ref{S0C}) that 
\begin{equation}
S^{(0)}(\phi_c)~=~-\frac{1}{2}\sum_\bs\;\bc(\bs)\cdot\bD(\bs,\s_2,\s_1)\cdot\bc(\bs)\;,\label{S0C_SUM}
\end{equation} 
with 
\begin{equation} 
\fl \bD~=~\bF^{-1}\bG\;,~~~~~~ \bG(\bs,\s_2,\s_1)~=~ \left(
\ba{ccc}
-J(\s_1)F_1^\prime(\bs,\s_1)&~~~~&J(\s_2)F_1^\prime(\bs,\s_2)\\ &&\\
-J(\s_1)F_2^\prime(\bs,\s_1)&~~~~&J(\s_2)F_2^\prime(\bs,\s_2) \ea
\right)\;. \label{D} 
\end{equation} 
Then we have 
\begin{eqnarray}
\fl \hK^{(0)}(\phi_2(\bx),\s_2;\phi_1(\bx),\s_1)&=&\prod_\bs\;K^{(0)}(\bs,c_2(\bs),\s_2;c_1(\bs),\s_1)\;,\nn\\
\fl K^{(0)}(\bs,c_2(\bs),\s_2;c_1(\bs),\s_1)&=&
A(\bs,\s_2,\s_1)\;\exp\left(-\frac{1}{2}\bc(\bs)\cdot\bD(\bs,\s_2,\s_1)\cdot\bc(\bs)\right)\;,
\label{K0} 
\end{eqnarray} 
where the normalisation factors $A(\bs,\s_2,\s_1)$
arise from the Gaussian integration over the fluctuations $\xi(\bx,t)$
of the field $\phi(\bx,t)$ about the classical solution. We have that
\begin{equation}
\fl A(\bs,\s_2,\s_1)~=~\int\;\cD\txi\;\exp\left(-\frac{1}{8\pi}\int_{\s_1}^{\s_2}\;d\s\:J(\s)\;
\left[\left(\frac{d\txi}{d\s}\right)^2~+~(\lambda(\bs,\s)+1)\txi^2\right]\right)\;,\label{A}
\end{equation} 
where 
\begin{eqnarray} 
\phi(\bx,\s)&=&\phi_c(\bx,\s)+\xi(\bx,\s)\;,\\\nn
\tphi(\bs,\s)&=&\tphi_c(\bs,\s)+\txi(\bs,\s), 
\end{eqnarray} 
and where generically we have defined the transform, $\tilde{f}(\bs,\s)$, of a function $f(\bx,\s)$ by 
\begin{equation} 
\tilde{f}(\bs,\s)~=~\int
d\bx\;f(\bx,\s)X(\bs,\bx)\;.  
\end{equation} 
The boundary conditions are 
\begin{equation}
\xi(\bx,\s_1) = \xi(\bx,\s_2) = 0~~\Longrightarrow~~ \txi(\bs,\s_1) =
\txi(\bs,\s_2) = 0,~~~~\forall~\bs\;. \label{XIBC} 
\end{equation} 
Then we have
\begin{equation}
\fl A(\bs,\s_2,\s_1)~\propto~\left(\det[\cL_\s(\bs)]\right)^{-1/2},~~~
\cL_\s(\bs)~=~-\frac{d}{d\s}J(\s)\frac{d}{d\s} + J(\s)(\lambda(\bs)+1).  
\end{equation} 
The determinant can be calculated by
diagonalising $\cL_t(\bs)$ on a basis of orthonormal eigenfunctions
which satisfy the boundary conditions on $\xi(\s)$ given in
Eq. (\ref{XIBC}). Whilst yielding the correct result this is not the
quickest way to compute $A(\bs,\s_2,\s_1)$. The Pauli-van-Vleck
formula tells us that 
\begin{equation} 
A~=~\prod_\bs A(\bs,\s_2,\s_1)~=~
\left(2\pi \left|\det\left[\frac{\pd^2
S^{(0)}(\phi_c)}{\pd\phi_1\pd\phi_2}\right]\right|\right)^{1/2}\;.
\end{equation} 
Using the expression for $S^{(0)}(\phi_c)$ in
Eq. (\ref{S0C_SUM}), we find 
\begin{equation}
A(\bs,\s_2,\s_1)~=~\sqrt{\frac{|\bD_{12}(\bs,\s_2,\s_1)|}{2\pi}}\;. \label{APvV}\end{equation} 

The Pauli-Van Vleck formula  can be derived by analytically continuing the
Euclidean time variable $t$ to Minkowski time $\tau$, by performing
the Wick rotation $t \to \tau = -it$. For this purpose we consider the
kernels $\hK$ and $K^{(0)}$ as functions of $t_i = t(\s_i)$ rather
than $\s_i$, and then $\hK(\phi_2,i\tau_2;\phi_1,i\tau_1)$ as defined by
Eq. (\ref{K_L}) is a unitary operator which means that 
\begin{equation}
\fl K^{(0)}(\bs,c_2,i\tau;c_1,i\tau)) =  \int
dc^\prime\;\left(K^{(0)}(\bs,c^\prime,i\tau^\prime;c_2,i\tau)\right)^*
K^{(0)}(\bs,c^\prime,i\tau^\prime;c_1,i\tau)~=~\d(c_2-c_1), 
\end{equation} 
for any $\tau^\prime$. From Eq. (\ref{K0}) 
\begin{equation}
\fl K^{(0)}(\bs,c_2,i\tau_2;c_1,i\tau_1) = 
A(\bs,i\tau_2,i\tau_1)\exp\left(-\frac{i}{2}\bc(\bs)
\cdot\bD^I(\bs,\tau_2,\tau_1)\cdot\bc(\bs)\right)\;,
\end{equation} 
where $\bD^I(\bs,\tau_2,\tau_1) = -i\bD(\bs,i\tau_2,i\tau_1)$ is
a real symmetric matrix.  Then 
{
\bacol
\begin{eqnarray} 
\fl |A(\bs,i\tau_2,i\tau_1)|^2\int & & 
dc^\prime\;\exp\left(\frac{i}{2}\left[2\bD^I_{12}(\bs,i\tau_2,i\tau_1)(c_2-c_1)c^\prime
+ \bD^I_{22}(\bs,i\tau_2,i\tau_1)(c_2^2-c_1^2)\right]\right) \nonumber \\ 
&&=\d(c_2-c_1).
\end{eqnarray} 
}
This equation determines $A(\bs,i\tau_2,i\tau_1)$ and, on
analytic continuation back to Euclidean time $t$ and re-expressing as a
function of $\s$, we find that $A(\bs,\s_2,\s_1)$ is given by
Eq. (\ref{APvV}).

The kernel $\hK^{(0)}(\phi_2(\bx),i\tau_2;\phi_1(\bx),i\tau_1)$
analytically continued to Minkowski time $\tau$ describes the time
evolution of the wave-function in the associated quantum mechanics
problem.  The Hamiltonian associated with
$K^{(0)}(\bs,c_2,t_2;c_1,t_1)$ is 
\begin{equation}
{\cal H}(\bs,c,t)~=~-\frac{1}{2}\frac{\pd^2}{\pd
c^2}~+~J^2(\s)(\lambda(\bs,\s)+1)\,c^2\;, 
\end{equation} 
where $\s \equiv \s(t)$
is defined by inverting Eq. (\ref{etime}). This Hamiltonian contains
an explicit time dependence and so the usual quantum mechanical
analysis becomes more general. The Euclidean version of the
Schr\"odinger equation for wave-function $\psi(\bs,c,t)$ is 
\begin{equation}
-\frac{\pd}{\pd t}\psi(\bs,c,t)~=~{\cal H}(\bs,c,t)\psi(\bs,c,t)\;.  
\end{equation}
This equation is also satisfied by
$K^{(0)}(\bs,c,t;c^\prime,t^\prime)$ regarded as a function of $(c,t)$
for fixed $(c^\prime,t^\prime)$. As remarked earlier, because the
Hamiltonian is explicitly $t$-dependent there are no stationary states
associated with ${\cal H}(\bs,c,t)$. However, in the limit that either $t \to
\infty$ or $t \to -\infty$ the kernel
$K^{(0)}(\bs,c,t;c^\prime,t^\prime)$ is a separable function of $c$
and $c^\prime$ except in one particular case. These properties will be
elucidated in the context of cylindrical interfaces discussed in the
next section. The connection between the grand partition function in
statistical mechanics and the related quantum mechanical formulation
is the one outlined above between the imaginary and real time
formalism (\cite{feyn}).

The final outcome for the contribution from quantum numbers $\bs$ to
the kernel for the free field theory in the layer, up to an irrelevant
factor, is 
\begin{equation}
\fl K^{(0)}(\bs,c_2,\s_2,c_1,\s_1)~=~\frac{1}{\sqrt{|H(\bs,\s_2,\s_1)|}}\:
\exp\left(-\frac{1}{2}\bc\cdot\bD(\bs,\s_2,\s_1)\cdot\bc\right),
\label{K0MEM} 
\end{equation} 
where, using Eqs. (\ref{F}) and (\ref{D}), we find
\begin{equation} \bD(\bs,\s_2,\s_1)~=~\frac{1}{H(\bs,\s_2,\s_1)} \left( \ba{cc}
W(\bs,\s_2,\s_1) & 1 \\ &\\ 1 & W(\bs,\s_1,\s_2) \ea
\right)\;. \label{Dmem} 
\end{equation} 
We have used the identity for the
Wronskian in Eq. (\ref{wronskian}), and have defined 
\begin{eqnarray}
W(\bs,\s_j,\s_i)&=&J(\s_i)[F_1(\bs,\s_j)F_2^\prime(\bs,\s_i)-F_1^\prime(\bs,\s_i)F_2(\bs,\s_j)]\;,
\nn\\
H(\bs,\s_j,\s_i)&=&F_1(\bs,\s_i)F_2(\bs,\s_j)-F_2(\bs,\s_i)F_1(\bs,\s_j)\;. \label{WandH}
\end{eqnarray}

\section{\label{cylinder} Concentric cylinders}
We now apply the formalism of the previous section to the case of two
concentric cylinders of length $L$ in the $z$-direction and radii
$r_1$ and $r_2$, respectively , with $r_1 < r_2$.  The separable
coordinates are $\bx = (\theta,z),\;\s = r$ and the Euclidean time
coordinate is $t = \log(r)$ (note, all coordinates are considered
dimensionless at this stage). Hence, $r \to 0(\infty)~\Rightarrow~t
\to -\infty(\infty)$. In what follows, we work with $r$ rather than
$t$ for simplicity. The volume measure is 
\begin{equation} 
dv = rdr d\theta dz
\Rightarrow J(r)=r\;, 
\end{equation} 
and Eq. (\ref{Xsx}) becomes 
\begin{equation}
-\left(\frac{1}{r^2}\frac{\pd^2}{\pd\theta^2}~+~\frac{\pd^2}{\pd
z^2}\right)X(\bs,\theta,z) ~=~\lambda(\bs,r)X(\bs,\theta,z)\;, 
\end{equation}
with solution 
\begin{equation} 
\ba{c} \ds
X(\bs,\theta,z)~=~\frac{1}{2\pi}e^{in\theta}\:e^{ipz}\;.\\[2mm] \ds
\bs = (n,p)\;,~n \in Z,~-\infty < p < \infty\;,\\[2mm] \ds
\lambda(\bs,r) = (n^2/r^2+p^2)\;.  
\ea 
\end{equation} 
Eq. (\ref{Tst}) is then
\begin{equation} 
\left[-\frac{d}{dr}r\frac{d}{dr}~+~\frac{n^2}{r}+(p^2+1)r\right]
T(\bs,r)~=~0\;.  
\end{equation} 
This is Bessel's modified equation \cite{grad}, and the two
required solutions are 
\begin{equation}
F_1(\bs,r)~=~I_n(Pr)\;,~~F_2(\bs,r)~=~K_n(Pr)\;.  
\end{equation} 
where $P^2 =
p^2+1$. We note that the Wronskian condition of Eq. (\ref{wronskian})
is satisfied by these solutions since 
\begin{equation}
Pr[I_n(Pr)K_n^\prime(Pr)-I_n(Pr)I_n^\prime(Pr)]~=~1\;. \label{WRONK}
\end{equation} 
Then, using Eqs. (\ref{Dmem}) and (\ref{WandH}), we find 
\begin{equation}
\bD(\bs,r_2,r_1)~=~\frac{1}{H_n(Pr_2,Pr_1)} \left( \ba{cc}
W_n(Pr_2,Pr_1) & 1 \\ &\\ 1 & W_n(Pr_1,Pr_2) \ea \right)\;,
\label{Dcyl} 
\end{equation} 
where 
\begin{eqnarray}
W_n(Pr_j,Pr_i)&=&Pr_i[I_n(Pr_j)K_n^\prime(Pr_i) -
I_n^\prime(Pr_i)K_n(Pr_j)]\;,\nn\\ H_n(Pr_j,Pr_i)&=&I_n(Pr_i)K_n(Pr_j)
- K_n(Pr_i)I_n(Pr_j)\;. \label{WnandHn} 
\end{eqnarray} 
The contribution from
quantum numbers $\bs=(n,p)$ to the kernel for the free field theory in
the layer, up to an irrelevant constant factor, is then 
\begin{equation}
\fl K^{(0)}(\bs,c_2,r_2,c_1,r_1)~=~\frac{1}{\sqrt{|H_n(Pr_2,Pr_1)|}}\:
\exp\left(-\frac{1}{2}\bc\cdot\bD(n,p,r_2,r_1)\cdot\bc\right). \label{K0CYL}
\end{equation}

\subsection{Asymptotic behaviour}
We use the definition of $K^{(0)}(\bs,c^\prime,r^\prime,c,r)$ in
Eq. (\ref{K0CYL}) and the asymptotic behaviour for the Bessel
functions given in Eq. (\ref{ASYM}) to derive the behaviour of
$K^{(0)}$ as $r \to 0$ and $r \to \infty$. Because we need to consider
the case when the Debye mass $m$ is zero we carry out the analysis
using dimensionful coordinates. This follows easily if we interpret
$p$ and $r$ as carrying dimension, with $P^2 = p^2+m^2$, and rescale
$\bD \to \beta\epsilon\bD$.

\subsubsection{\boldmath $r^\prime \to \infty$}
In the limit $r^\prime \to \infty$ ($t^\prime \to \infty$) the natural
boundary condition for the scalar field is $\phi = 0$, which
corresponds to $c^\prime = 0$, and we impose this condition from now
on. For the various cases we find

\bit
\item[]{\boldmath $p = m = 0:$} 
\begin{eqnarray}
\fl K^{(0)}(n=0,0,c^\prime=0,r^\prime;c,r)~&\sim&~
\frac{1}{\sqrt{\log(r^\prime/r)}}\exp\left(-\frac{\bete}{2}\frac{c^2}{\log(r^\prime/r)}\right)\nonumber \\
&=&\frac{1}{\sqrt{t^\prime-t}}\exp\left(-\frac{\bete}{2}\frac{c^2}{(t^\prime-t)}\right)\;. \label{ASK0}
\end{eqnarray} 
This is the free particle kernel for Euclidean time
(\cite{feyn}). It is the one case where the kernel is not separable in
the $c,c^\prime$ variables. The important feature of this result for
the charge neutrality condition is that 
\begin{equation} 
\frac{\pd}{\pd
c}\log(K^{(0)})~=~-\bete\frac{c}{(t^\prime-t)}~\to~0~~\mbox{as}~~t^\prime~\to~\infty\;. \label{deriv_s=0a}
\end{equation} 
Also 
\begin{equation} 
\fl K^{(0)}(n>0,0,c^\prime=0,r^\prime;c,r)~\sim~
\sqrt{2n}
\exp\left(-\frac{n}{2}\log{r^\prime/r}\right)\exp\left(-\frac{\bete}{2}nc^2\right)\;.
\label{ASK1}
\end{equation} 
This is the harmonic oscillator ground-state in $c$ with
associated energy $E_0 = n/2$. The prefactor contains the correct
(Euclidean) time-dependent factor $\exp(-n/2(t^\prime-t))$.

\item[]{\boldmath $P = \sqrt{p^2+m^2} > 0:$}

We define the function $V_n(z)$ by 
\begin{equation}
V_n(z)~=~-\frac{zK_n^\prime(z)}{K_n(z)}\;.\label{ASK2} 
\end{equation} 
Then 
\begin{eqnarray}
\fl K^{(0)}(n,p,c^\prime=0,r^\prime;c,r) \nn\\
\lo\sim (2\pi Pr^\prime)^{1/4}\exp\left(-\frac{1}{2}Pr^\prime\right)
\frac{1}{\sqrt{K_n(Pr)}}\exp\left(-\frac{\bete}{2}V_n(Pr)c^2\right).\label{ASK3}
\end{eqnarray} 
\eit 
The related Schr\"odinger equation, which has a
time-dependent Hamiltonian, satisfied by
$K^{(0)}(\bs,c^\prime,r^\prime;c,r)$ considered as a function of $c$
and $t = \log(r)$, is 
\begin{equation} 
-\frac{\pd}{\pd t}\psi(\bs,c,t)~=~
\left(-\frac{1}{2}\frac{\pd^2}{\pd c^2} +
\frac{1}{2}\left(P^2e^{2t}+n^2){c}^2\right)\right) \psi(\bs,c,t)\;.
\end{equation} 
It can be verified that the different forms listed above in the
limit $t^\prime \to \infty$ do, indeed, satisfy this equation.

\subsubsection{\boldmath $r \to 0$}

In the limit $r \to 0~(t \to -\infty)$, for the various cases, we find

\bit

\item[]{\boldmath $p = m = 0:$} 
\begin{eqnarray}
 K^{(0)}(n=0,0,c^\prime,r^\prime;c,r)~&\sim&
\frac{1}{\sqrt{\log(r^\prime/r)}}\exp\left(-\frac{\bete}{2}\frac{(c-c^\prime)^2}{\log(r^\prime/r)}\right)\nonumber \\
&=&\frac{1}{\sqrt{t^\prime-t}}\exp\left(-\frac{\bete}{2}\frac{(c-c^\prime)^2}{(t^\prime-t)}\right)\;.
\label{ASK4}
\end{eqnarray} 
This is the free particle kernel for Euclidean time
(\cite{feyn}). As before 
\begin{equation} 
\frac{\pd}{\pd
c^\prime}\log(K^{(0)})~\to~0~~\mbox{as}~~t~\to~-\infty\;. \label{deriv_s=0b}
\end{equation} 
which we shall show ensures charge neutrality. Also
  
\begin{equation}
\fl  K^{(0)}(n>0,0,c^\prime,r^\prime;c,r)~\sim~
\sqrt{2n}\exp\left(-\frac{n}{2}\log{r^\prime/r}\right)\exp\left(-\frac{\bete}{2}n(c^2+{c^\prime}^2)\right)\;.
\label{ASK5}
\end{equation} 
This is the harmonic oscillator ground state in both $c$ and
$c^\prime$ with associated energy $E_0 = n/2$.  The prefactor contains
the correct (Euclidean) time-dependent factor
$\exp(-n/2(t^\prime-t))$.

\item[]{\boldmath $P = \sqrt{p^2+m^2} > 0:$}

We define the function $U_n(z)$ by 
\begin{equation}
U_n(z)~=~\frac{zI_n^\prime(z)}{I_n(z)}\;. \label{ASK6} 
\end{equation} 
Then 
\begin{equation}
\fl K^{(0)}(n=0,p,c^\prime,r^\prime;c,r) \sim \frac{1}{\sqrt{-\log(r)}}
\frac{1}{\sqrt{I_0(Pr^\prime)}}\exp\left(-\frac{\bete}{2}U_0(Pr^\prime){c^\prime}^2\right),
\label{ASK7} 
\end{equation} 
\begin{eqnarray}
\fl K^{(0)}(n>0,p,c^\prime,r^\prime;c,r) \sim \nn\\
\lo \exp\left(\frac{1}{2}n\log(Pr)\right)
\exp\left(-\frac{\bete}{2}nc^2\right)\frac{1}{\sqrt{I_n(Pr^\prime)}}
\exp\left(-\frac{\bete}{2}U_n(Pr^\prime){c^\prime}^2\right). \label{ASK8}
\end{eqnarray} 
\eit
As before these asymptotic forms satisfy the related Schr\"odinger equation.

\subsection{\label{cylinder_membrane} The cylindrical membrane}
The i-th layer of a system of $N$ concentric cylindrical layers has
volume $V_i$ and is bounded by the cylindrical surfaces $S_i$ and
$S_{i+1}$, where $S_0$ is the innermost surface and $S_N$ is the
outermost.  The Debye mass and dielectric constant associated with the
bulk reservoir connected to the $i$-th layer are denoted by
$m_i,\epsilon_i$, respectively. The tube is of length $L$ in the
$z$-direction which is parallel to the symmetry axis of the
cylinders. The contribution to the grand partition function from this
system is the convolution 
\begin{equation} 
\Xi_M~=~\int \cD
c\;\prod_\bs\prod_{i=0}^{N-1}\;
K_i(\bs,c_{i+1}(\bs),r_{i+1};c_i(\bs),r_i)\;,\label{XIM} 
\end{equation} where
$\bs=(n,p)$ as before and the field measure is 
\begin{equation} 
\cD c~=~\prod_\bs
\prod_{i=0}^N\;dc_i(\bs)\;, 
\end{equation} 
with boundary condition
$c_N(\bs)=0~\forall~\bs$.

The grand partition function for the whole system including the bulk
reservoirs to which the different layers connect is 
\begin{equation}
\Xi~=~\frac{\Xi_M}{\Xi_B}~~~~~~\Xi_B~=~\prod_{i=0}^{N-1}\;\Xi_{B_i}(V_i)\;,
\end{equation} 
where $\Xi_{B_i}(V_i)$ is the bulk grand partition function for
$i-th$ layer of volume $V_i$. This can be calculated using the bulk
action defined in Eq. (\ref{S_B}) with chemical potential $\mu_i$ and
dielectric constant $\e_i$ for a torus of volume $V_i$.

The free energy is then $F =- k_BT\log\Xi$ and the forces acting on the
interfaces and the stability of the system can be deduced from
$F$. From Eq. (\ref{S0}) and the earlier discussion the perturbation
theory for $F$ can be developed as a loop expansion with expansion
parameters $g_i = m_il_B$ where $m_i$ is the Debye mass of the $i$-th
layer and $l_B$ is the Bjerrum length. The expansion is a cumulant
expansion about the quadratic, or free, field theory which is
described by the quadratic approximation to the grand partition
function, $\Xi^{(0)}$ of the system 
\begin{eqnarray}
\Xi^{(0)}&=&\frac{\Xi^{(0)}_M}{\Xi^{(0)}_B}\;,\nn\\
\Xi^{(0)}_M&=&~\int \cD c\;\prod_\bs\prod_{i=0}^{N-1}\;
K^{(0)}_i(\bs,c_{i+1}(\bs),r_{i+1};c_i(\bs),r_i)\;,
\label{XI0}
\end{eqnarray} 
and with $\Xi^{(0)}_B$ defined in terms of the free-field action
as described just above. Each term in the product over $\bs$ on the
RHS of Eq. (\ref{XI0}) has an exponent which is a quadratic form in
the interface field variables $\bc(\bs)$, where the notation
$\bc(\bs)$ has been used to signify the vector of all the $c_i(\bs)~~0
\le i \le N$ associated with the given set of quantum numbers $\bs$.
The integral over the boundary fields with respect to the measure $\cD
c$ is therefore Gaussian, and can be done exactly. The free energy
$F^{(0)} = -k_BT\log\Xi^{(0)}$ contains the ideal gas contribution and
this one-loop term. The one-loop term consists of a contribution from
the normalisation factors of the $K^{(0)}_i$, and from the determinant
of the matrix defining the quadratic form in the exponent which arises
from the Gaussian integration over the boundary field values. The
Casimir forces acting on the system are determined by the one-loop
contribution. In \cite{deho} the attractive Casimir forces acting
between the faces of a planar soap film were discussed and derived in
this manner, and the contribution to two-loop order in the cumulant
expansion of the interaction $\Delta {\cal S}$ defined in
Eq. (\ref{DS}) for the planar film was presented in
\cite{dehotloop}. In general, the loop perturbation theory can be
carried out in the same way for any symmetrical layered electrolytic
system such as that constructed from concentric cylinders or
spheres. This perturbation theory will be pursued in a future
publication. In the next section we analyse the case of a thin
cylindrical membrane for which $N=3$.

\section{\label{cylinder_casimir}The Casimir force for a dielectric tube}
\begin{figure}
\begin{center}
\includegraphics[width=.5\textwidth]{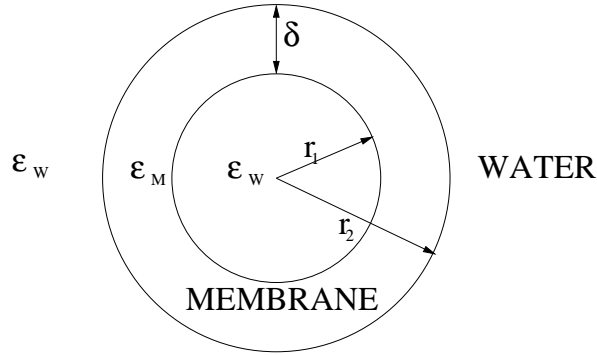}
\end{center}
\caption{Horizontal cross section through idealised tubule
configuration shown are dimensions and dielectric permittivities}
\label{tube1}
\end{figure}

In Figure \ref{tube1} the cross-section of a tube of inner radius $R$
formed from a membrane of thickness $\d$ is shown, with radii for the
boundary surfaces defined to be 
\begin{equation} r_0 = 0,~~r_1 =
R-\frac{\d}{2},~~r_2 = R+\frac{\d}{2},~~r_3 = \infty\;, 
\end{equation} with $\d\ll R$.

\begin{figure}
\begin{center}
\includegraphics[width=.5\textwidth]{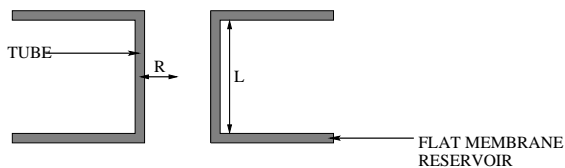}
\end{center}
\caption{Vertical cross section through idealised tubule configuration
showing tubule of length $L$ and radius $R$ bridging two flat
bilayers}
\label{tube2}
\end{figure}
In this paper we concentrate on the Casimir force acting on the tube
described above and shown in Figure \ref{tube2} in which the
electrolyte densities, and hence the Debye masses, are zero in all
three layers, the membrane is of fixed thickness $\d$ with dielectric
constant $\epsilon_2=\epsilon_M = 2\epsilon_{vac}$, and the inner and
outer layers are filled with water so that
$\epsilon_0=\epsilon_3=\epsilon_W = 80\epsilon_{vac}$. The Casimir
force is thus due purely to the discontinuity in the dielectric
constants at the membrane surfaces and is a function of the radius $R$
of the inner cylindrical layer. We shall show that the Casimir force
in this case is attractive, tending to collapse the tube. The tube is
of length $L$ which we assume is large on any relevant scale, and that
there is a reservoir of membrane so that the tube radius $R$ can
change without the membrane needing to stretch. For example, the
system can be thought of as made from a flat sheet of membrane on to
which the tube connects and which acts as a reservoir of membrane as
the tube expands or contracts as shown in Figure
(\ref{tube2}). Alternatively, the membrane can be
folded at one end of the tube and act as a reservoir. Both scenarios
are possible in biological systems where the membrane is a lipid
bilayer, although for the latter it is difficult to calculate the free
energy of a given volume of lipid in the reservoir so presents a
problem of normalisation. However, it should be emphasised that this
picture may nevertheless be an important feature of the stability of
lipid tubules and needs further analysis. In either case, because we
assume that the membrane is not stretched as the area of the tube
increases, there is no elastic energy stored in the tube except that
due to the curvature and the surface area of the system is 
constant.
In what follows we shall assume a reservoir of
flat membrane as shown in Figure (\ref{tube2}).

As an intermediate step we define the grand partition function,
$\Xi^{(0)}_{MW}(R,\d)$, for the membrane normalised by that of an
equivalent water filled region, and its associated free energy
$F^{(0)}_{MW}(R,\d)$ by 
\begin{equation}
\fl \Xi^{(0)}_{MW}(R,\d)~=~\frac{\Xi^{(0)}_M(R,\d)}{\Xi^{(0)}_W(R=\infty,\d)}\;,~~~~
F^{(0)}_{MW}(R,\d)~=~-k_BT\log\Xi^{(0)}_{MW}(R,\d)\;,\label{XI0MWdef}
\end{equation} 
where $\Xi^{(0)}_M$ is given by Eq. (\ref{XI0}), and where
$\Xi^{(0)}_W$ is the grand partition function of a system filled with
water only: $\Xi^{(0)}_W = \Xi^{(0)}_M$ for $\delta = 0$. For the
grand ensemble with a reservoir consisting of a flat membrane of the
same thickness, shown in Figure \ref{tube1}, we must subtract the free
energy of a flat membrane of equivalent area to the tube. We then find
that the free energy appropriate to calculate the Casimir force due to
the tube geometry is 
\begin{equation}
F_C(R,\delta)~=~F^{(0)}_{MW}(R,\delta)~-~2\pi RL\,F_\infty(\delta)\;,
\label{Free} \end{equation} 
where 
\begin{equation} 
F_\infty(\delta)~=~\lim_{R \to
\infty}\frac{F^{(0)}_{MW}(R,\delta)}{2\pi RL}\;, \label{Finf} 
\end{equation}
where $F_\infty(\delta)$ is the free energy per unit area of flat
membrane of thickness $\delta$. Using the expression for
$\Xi^{(0)}_M(R,\d)$ in Eq. (\ref{XI0}) and the asymptotic expressions
for $K^{(0)}$ as $r \to 0$ in Eqs. (\ref{ASK5}) and (\ref{ASK6}) it
can be seen that the dependence on $c_0(\bs)~$, which are the
coefficients determining the boundary field value $\phi_0(\bx)$ at
$r=0$, cancels out $\;\forall \bs\;$ between numerator and denominator
in Eq. (\ref{XI0MWdef}). This is independent of whether the integrals
over the $c_0(\bs)$ are done or not; it is a consequence only of
separability in the limit $r \to 0$ or, in the case $n=0$, that
$K^{(0)}$ becomes independent of the $c_0(\bs)$. Thus we may set
$c_0(\bs) = 0~~\forall \bs$ in what follows and omit the integrals
over the $c_0(\bs)$.

The difference between the numerator and denominator in
$\Xi^{(0)}_{MW}(R,\d)$ in Eq. (\ref{XI0MWdef} is then due to the
different contributions of the membrane layer between radii
$r_1=R-\d/2$ and $r_2=R+\d/2$ which has dielectric constant
$\epsilon_M$ in the numerator but $\epsilon_W$ in the denominator. We
then have 
\begin{equation}
\fl \Xi^{(0)}_{MW}(R,\d)=\int \prod_\bs
dc_1(\bs)dc_2(\bs)\;Q^{(0)}(\bs,c_2(\bs),r_2;,c_1(\bs),r_1)\:
K^{(0)}_1(\bs,c_2(\bs),r_2;,c_1(\bs),r_1), \label{XI0MWeq}
\end{equation}
where
\begin{equation}
Q^{(0)}(\bs,c_2,r_2;,c_1,r_1)=\lim_{{r_3 \to \infty} \atop {r_0 \to
0}}
\frac{K_0(\bs,0,r_3;c_2,r_2)K_2(\bs,c_1,r_1;0,r_0)}{K(\bs,0,r_3;0,r_0)}\;,
\label{Q} 
\end{equation} 
where in $Q^{(0)}$ the dielectric constant in the
denominator kernel is $\epsilon_W$. From the asymptotic expressions in
Eqs. (\ref{ASK0}-\ref{ASK8}) we find that $Q^{(0)}$ has a simple form
for $(n,p) \ne (0,0)$ ( c.f. $\bs = (n,p)$) 
\begin{eqnarray}
\fl &&Q^{(0)}(\bs,c_2,r_2;,c_1,r_1)\nonumber \\
\fl&=&\left[\frac{\bete_W}{2\pi
|K_n(pr_2)I_n(pr_1)|}\right]^{1/2}
\exp\left(\frac{-\bete_W}{2}V_n(pr_2)c_2^2\right)\:\exp\left(-\frac{\bete_W}{2}U_n(pr_1)c_1^2\right)\;,
\label{Q0} \end{eqnarray} 
where $U_n,V_n$ are defined in Eqs. (\ref{ASK7}) and
(\ref{ASK2}), respectively. Note, that all Debye masses are zero
here. For large argument $U_n, V_n \to 1$, and so both Gaussian forms
are convergent and integrable.  $Q^{(0)}$ takes the form of a
normalised product of a generalisation of harmonic oscillator
ground state wave-functions in $c_1$ and $c_2$, and satisfies the
appropriate Schr\"odinger equations in these variables. For large $R$
and $n > 0$ these functions become the usual ground-state oscillator
wave-functions with $m\omega = \bete_Wn$, which agrees with the
analysis of the planar film of \cite{deho}.

For $(n,p) = (0,0)$ we find 
\begin{equation}
\fl Q^{(0)}(\bs=\bzero,c_2,r_2;,c_1,r_1)~=~\lim_{{r_3 \to \infty} \atop
{r_0 \to 0}} \left[\frac{\bete_W}{2\pi}\frac{t_{30}}{t_{32}t_{10}}\right]^{1/2}
\exp\left(-\frac{\bete_W}{2}\frac{c_2^2}{t_{32}}\right)\exp\left(-\frac{\bete_W}{2}\frac{c_1^2}{t_{10}}\right),
\label{Qs=0}
\end{equation} 
where $t_i = \log r_i$ and $t_{ij} = t_i-t_j$.  Also, we have that
\begin{equation} 
\fl K^{(0)}_1(\bs=\bzero,c_2(\bs),r_2;,c_1(\bs),r_1)~=~
\left[\frac{\bete_M}{2\pi}\frac{1}{t_{21}}\right]^{1/2}
\exp\left(-\frac{\bete_W}{2}\frac{(c_1-c_2)^2}{t_{21}}\right)\;,
\end{equation} 
and so we find the contribution from $\bs = \bzero$ mode to the
partition function is 
\begin{equation}
\Xi^{(0)}_{MW}(\bs=\bzero,R,\d)~=~~\lim_{{r_3 \to \infty} \atop {r_0
\to 0}} \left[\frac{\e_W\e_M t_{30}} {\e_W(\e_W-\e_M)t_{21} +
{\e_W\e_M}t_{30}}\right]^{1/2}~=~1\;.  
\end{equation} 
Thus, the zero mode does
not contribute to the free energy of the membrane. However, it is the
relevant mode to show that the charge neutrality condition holds. The
total charge operator $\Sigma$ is given by 
\begin{equation} \Sigma~=~\e_W\int d\bx
E(\bx,r_2+\eta) - \e_W\int d\bx E(\bx,r_1-\eta)\;,\label{charge} 
\end{equation}
where $E(\bx,r)$ is the radial component of the electric field and
$\eta$ is a small positive length; thus, we measure the field in the
water just outside the membrane surfaces. We have that 
\begin{equation}
E(\bx,t)~=~-\frac{\pd}{\pd r}\la \phi(\bx,r) \ra ~=~
-\frac{1}{J(t)}\la \frac{\pd}{\pd t}\phi(\bx,t) \ra
~=~-\frac{1}{J(r(t)\bete_W}\la \pi(\bx,t) \ra\;, 
\end{equation} 
now considering
$E$ and $\phi$ as functions of $t=\log(r)$. Here $\pi(\bx,t)$ is the
momentum operator conjugate to $\phi(\bx,t)$ and is given by
$\pi(\bx,t) = \bete \dot\phi(\bx,t)$ using standard theory. The
Schr\"odinger representation of $\pi(\bx,t)$ \cite{dehocon} then gives
\begin{equation} E(\bx,t)~=~\frac{1}{J(t)\bete}\la \phi,t|\;\frac{\delta}{\delta
\phi(\bx)}\;|\phi,t \ra\;.  
\end{equation} 
The contribution to $\la \Sigma \ra$
from the integral over the surface at $r=r_1$ is then 
\begin{equation} \la
\Sigma_1 \ra~=~\frac{1}{J(t)\beta}\la \tphi,t|\;\frac{\delta}{\delta
\tphi(\bzero)}\;|\tphi,t \ra\;, 
\end{equation} 
where $\tphi(\bzero)$ is the
zero-mode field. Using Eqs. (\ref{deriv_s=0a}) and (\ref{Qs=0}) we
then find 
\begin{equation} \la \Sigma_1 \ra~=~\lim_{t_0 \to
-\infty}\frac{\e_W}{J(t_1)}\frac{\la
\tphi,t_1|\;\tphi(\bzero)\;|\tphi,t_1 \ra}{t_1-t_0} ~=~0\;.  
\end{equation} 
A similar result holds for $\la \Sigma_2 \ra$, the contribution to $\la
\Sigma \ra$ from the surface integral at $r=r_2$; thus we find that
$\la \Sigma \ra = 0$. This analysis can be repeated to show that all
moments of $\Sigma$ vanish: $\la \Sigma^n \ra = 0,~\forall n > 0$; it
is this condition that ensures charge neutrality of the system.

We now calculate the free energy $F^{(0)}$ by summing over all $\bs =
(n,p)$ mode contributions. From the previous section we find 
\begin{eqnarray}
\fl K^{(0)}_1(\bs,c_2(\bs),r_2;,c_1(\bs),r_1) \nn\\
\lo= \left[\frac{\bete_M}{2\pi |H_n(pr_2,pr_1)|}\right]^{1/2} 
\exp\left(-\frac{\bete_M}{2}\bc(\bs)\cdot \bD(\bs,r_2,r_1) \cdot \bc(\bs)\right), \label{K012} 
\end{eqnarray} 
where as before $\bc = (c_1,c_2)$
and 
\begin{equation} \bD(\bs,r_2,r_1)~=~\frac{1}{H_n(pr_2,pr_1)} \left( \ba{cc}
W_n(pr_2,pr_1) & 1 \\ &\\ 1 & W_n(pr_1,pr_2) \ea \right)\;,
\label{Dcyl0} 
\end{equation} 
and $W_n$ and $H_n$ are defined in
Eq. (\ref{WnandHn}). Using Eqs. (\ref{XI0MWeq}), (\ref{Q0}), and
(\ref{K012}) we find 
\begin{eqnarray}
\Xi^{(0)}_{MW}(R,\d)&=& \nn\\ \int
\prod_\bs\frac{1}{2\pi} &dc_1&(\bs)dc_2(\bs)
\left[\frac{\bete_W}{|K_n(pr_2)I_n(pr_1)|}\right]^{1/2}
\left[\frac{\bete_M}{|H_n(pr_2,pr_1)|}\right]^{1/2}\nn\\
&&\exp\left(-\frac{\beta}{2}\bc(\bs)\cdot
\bE_M(\bs,r_2,r_1)\cdot\bc(\bs)\right)\;,\label{XI0a} 
\end{eqnarray} 
where 
\begin{equation}
\bE_M(\bs,r_2,r_1)~=~\e_M\bD(\bs,r_2,r_1)+\e_W\bX(\bs,r_2,r_1)\;, 
\end{equation}
and 
\begin{equation} \bX(\bs,r_2,r_1) = \mbox{diag}[U_n(pr_1),\; V_n(pr_2)]\;,
\label{X} 
\end{equation} 
is a $2 \times 2$ diagonal matrix. A more useful
alternative expression is 
\begin{equation}
\Xi^{(0)}_{MW}(R,\d)~=~\left(\frac{\e_M}{\e_W}\right)^{1/2}
\prod_\bs\frac{\displaystyle\int
d\bc(\bs)\exp\left(-\frac{\beta}{2}\bc(\bs)\cdot
\bE_M(\bs,r_2,r_1)\cdot\bc(\bs)\right)} {\displaystyle\int
d\bc(\bs)\exp\left(-\frac{\beta}{2}\bc(\bs)\cdot
\bE_W(\bs,r_2,r_1)\cdot\bc(\bs)\right)}\;,
\label{XI0b}
\end{equation} 
where 
\begin{equation}
\bE_W(\bs,r_2,r_1)~=~\e_W(\bD(\bs,r_2,r_1)+\bX(\bs,r_2,r_1))\;.  
\end{equation}
The denominator is the contribution from a pure water-filled
system. The two expressions for $\Xi^{(0)}_{MW}(R,\d)$ are the same
since using the Wronskian identity, Eq. (\ref{wronskian}), we find
\begin{equation}
\fl \det[\bD(\bs,r_2,r_1)+\bX(\bs,r_2,r_1)]~=~-[K_n(pr_2)I_n(pr_1)H_n(pr_2,pr_1)]^{-1}.
\end{equation} 
Then we have 
\begin{eqnarray}
\fl \Xi^{(0)}_{MW}(R,\d)&=&\frac{1}{2}\left(\frac{\e_M}{\e_W}\right)^{1/2}\prod_\bs
\det[\bB(\bs,r_2,r_1)]\;,\nn\\ \fl \bB(\bs,r_2,r_1)&=&\left[1 +
(\bX(\bs,r_2,r_1)+\bD(\bs,r_2,r_1))^{-1}\bX(\bs,r_2,r_1)\frac{(\e_W-\e_M)}{\e_M}\right]\;,
\label{XI0det}
\end{eqnarray} 
and the total free energy of the tube is 
\begin{eqnarray}
F_C(R,\d)&=&F^{(0)}_{MW}(R,\d) - F^{(0)}_f(\d)\;,\nn\\
F^{(0)}_{MW}(R,\d)&=&-\half
k_BT\log\left(\frac{\epsilon_M}{4\epsilon_W}\right) + L\sum_n\int
\frac{dp}{2\pi}\; F^{(0)}_{MW}(\bs,R,\d)\;,\nn\\
F^{(0)}_{MW}(\bs,R,\d)&=&-k_BT\log(\det[\bB(\bs,r_2,r_1)])\;.
\label{F0}
\end{eqnarray} 
The required free energy $F_C(R,\delta)$ is then given by
Eqs. (\ref{XI0MWdef}) and (\ref{Free}).  Using Eqs. (\ref{WRONK}) and
(\ref{WandH}) it can be verified, as expected, that
$F^{(0)}_{MW}(R,\d=0) = 0$. From Eq. (\ref{XI0b}) we find also that
$F^{(0)}_{MW}(R,\d) = 0$ when $\epsilon_W=\epsilon_M$ .

\section{\label{casimir_eval}Evaluation of the Casimir energy}
In this section we evaluate the Casimir energy for the dielectric tube
as a function of the inner radius $R$ formed from a membrane of fixed
thickness $\d$ and dielectric constant $\epsilon_M$. The regions
interior and exterior to the tube are water-filled with dielectric
constant $\epsilon_W$.  The cross-section of the tube is shown in
Figure \ref{tube2} and the length of the tube is aligned along the
$z$-axis.

The result for the free energy of the tube, $F_C(R,\d)$, is given in
Eqs. (\ref{F0}) as a sum over the mode number $n$ and an integral over
the wave-vector, $p$, in the $z$-direction. We evaluate the sum and
integral numerically and present results in the next section for
various values of $\d$.  However, as is usual in many cases, the
calculation is dominated by the Ultra-Violet (UV) properties of the
integrand and an UV cut-off must be imposed to achieve a finite
result. We examine the UV properties of the integral and calculate the
leading divergent contributions analytically. These divergent
contributions, which are regulated by the UV cut-off, agree with the
prediction for them obtained from the full numerical calculation. We
also verify that the $R \to \infty$ limit of $F^{(0)}_{MW}(R,\d)$
agrees with the result in the planar film case for a film of thickness
$\d$. It is convenient to define the following constants which encode
the dielectric properties of the system 
\begin{equation}
\Delta=\frac{\epsilon_W-\epsilon_M}{\epsilon_W+\epsilon_M}\;,~~~~
\gamma=\frac{\epsilon_W}{\epsilon_M}-1~=~\frac{2\Delta}{1-\Delta}\;.
\end{equation} 
After some algebra and use of the Wronskian identity
Eq. (\ref{WRONK}) we find 
\begin{equation}
\fl (\bD(\bs,r_2,r_1)+\bX(\bs,r_2,r_1))^{-1}~=~ \left( \ba{cc}
I_n(pr_1)K_n(pr_1) & I_n(pr_1)K_n(pr_2) \\ &\\ I_n(pr_1)K_n(pr_2) &
I_n(pr_2)K_n(pr_2) \ea \right). \label{DXinv} 
\end{equation} 
Using the
expression for the diagonal matrix $\bX$ in Eq. (\ref{X}) we find that
$\bB(\bs,r_2,r_1)$, whose determinant is required for the evaluation of
$\Xi^{(0)}$ in Eq. (\ref{XI0det}), is 
\begin{equation} 
\fl \bB(\bs,r_2,r_1)~=~ \left(
\ba{cc} 1+\gamma I^\prime_n(pr_1)K_n(pr_1) &
-I_n(pr_1)K^\prime_n(pr_2) \\ &\\ I^\prime_n(pr_1)K_n(pr_2) & 1-\gamma
I_n(pr_2)K^\prime_n(pr_2) \ea \right)\;. \label{B} 
\end{equation} 
The important
feature of $\bB$ is that the on-diagonal elements are the separate
contributions from the surfaces at $r = r_1$ and $r = r_2$ and the
off-diagonal terms are the contribution from the interaction between
the surfaces. In particular, it will be shown later in this section
that the off-diagonal elements fall off exponentially with the surface
separation $\delta$ like $\exp(-2\sqrt(p^2+m^2)\delta)$. This fact has
the consequences that the inter-surface interaction becomes negligible
for large separations or for large wave-vector $p$. The corollary is
that any Ultra-Violet divergences, which are due to large $p$
behaviour of the integrand, arise solely from the separate surface
contributions and that the inter-surface interaction gives a UV-finite
contribution. We therefore explicitly separate the terms in
$F^{(0)}_{MW}(R,\delta)$ into these respective contributions. We have
\begin{equation} 
\fl \frac{F^{(0)}_{MW}(R,\delta)}{Lk_BT}~=~\frac{1}{r_1}g(\L
r_1,\Delta) + \frac{1}{r_2}g(\L r_2,-\Delta) + h(r_1,r_2,\L,\Delta) +
m(\L,\Delta)\;,\label{Fghm} 
\end{equation} where 
\begin{eqnarray}
\fl g(x,\Delta)~&&=~\frac{1}{2\pi}\sum_n\int_0^x du \log[1+\Delta
(I_n(u)K_n(u))^\prime]\;,\nn\\ 
\fl h(r_1,r_2,\L,\Delta)~&&=~ \nn\\
\fl \frac{1}{2}\int_0^\L\frac{dk}{\pi}\sum_n&&\log\left[1+
\frac{\displaystyle
4\Delta^2p^2r_1r_2I^\prime_n(pr_1)I_n(pr_1)K^\prime_n(pr_2)K_n(pr_2)}
{\displaystyle(1+\Delta pr_1(I_n(pr_1)K_n(pr_1)^\prime)((1-\Delta
pr_2(I_n(pr_2)K_n(pr_2)^\prime)}\right]\;,\nn\\\fl
m(\L,\Delta)~&&=~-\frac{1}{2}\int_0^\L\frac{dk}{\pi}
\sum_n\log(1-\Delta^2)\;.\label{Ffns}
\end{eqnarray} 
The contribution $Lk_BTg(\L r,\Delta)/R$ is the free energy of
an isolated cylinder of length $L$, radius $R$ and dielectric constant
$\epsilon_M$ in a medium of dielectric constant $\epsilon_W$. Thus the
first two terms in Eq. (\ref{Fghm}) are the respective separate
contributions of the inner and outer cylindrical regions that form the
layer of thickness $\delta=r_2-r_1$; the term
$Lk_BTh(r_1,r_2,\Lambda,\Delta)$ is the contribution from the
interaction between the cylinders.  As expected, the function
$g(x,\Delta)$ diverges as $x \to \infty$ and so this term in the free
energy must be regulated by taking a finite non-zero cut-off $\L =
\pi/a$, where $a$ is the UV cut-off length. Viewed as a Taylor
expansion in $\Delta$ we find that the $O(\Delta)$ term of $g$ is
independent of $r$ and so in the free energy the contributions
proportional to $\Delta$ cancel. This to be expected on physical
grounds since by examining the limit of a diffuse system one can see
that any term proportional to $\Delta$ must be a self energy term
\cite{inprep}. The term of order $\Delta^2$ of $g$ can be evaluated
using Bessel function summation theorems  \cite{grad}.
This term is given by 
\begin{equation}
g_2(x,\Delta)~=~-\frac{\Delta^2}{4\pi}\int_0^x du\;
u^2\sum_n{[I_n(u)K_n(u)]^\prime}^2\;.\label{g2} 
\end{equation} 
We define
$R^*(r_1,r_2,\phi) = \sqrt{r_1^2+r_2^2-2r_1r_2\cos(\phi)}~$ and from
\cite{grad} we have 
\begin{equation} 
K_0(uR^*(r_1,r_2,\phi))~=~\sum_n
I_n(ur_1)K_n(ur_2)e^{in\phi}\;.\label{grad} 
\end{equation} 
Then 
\begin{eqnarray}
K_0(2uR^*(1,1,\phi))&=&\sum_nI_n(u)K_n(u)\;e^{in\phi}~~~~\Longrightarrow\nn\\
2\sin(\phi/2)\,K_0^\prime(2u\sin(\phi/2))&=&\sum_n\;[I_n(u)K_n(u)]^\prime\;e^{in\phi}\;.
\end{eqnarray} 
Thus 
\begin{equation}
\sum_n{[I_n(u)K_n(u)]^\prime}^2~=~\frac{2}{\pi}\int_0^{2\pi}K_1^2(2u\sin(\phi/2))\sin^2(\phi/2)\;,
\end{equation} 
where the Bessel function identity $K_0^\prime(u) = -K_1(u)$ has
been used. By substitution into Eq. (\ref{g2}) and careful
manipulation of the double integral we find that 
\begin{equation}
\fl g(x,\Delta)~=~-\frac{1}{256}\Delta^2\left[6\log(x) +30\log 2 +
6\gamma-11\right]~+~O(\Delta^4)+O(1/x)\;.
\label{gD2}
\end{equation} 
Similarly, the finite contribution $h(r_1,r_2,\L,\Delta)$ can be
expanded and the $O(\Delta^2)$ term evaluated.  We have 
\begin{equation}
\fl h_2(r_1,r_2,\L,\Delta)~=~ \frac{2}{\pi}\Delta^2\int_0^\L
dp\;r_1r_2\sum_n\;p^2I^\prime_n(pr_1)I_n(pr_1)K^\prime_n(pr_2)K_n(pr_2)\;.
\label{h2}
\end{equation} 
Using Eq. (\ref{grad}) we have 
\begin{eqnarray} 
\fl \sum_n p^2I^\prime_n(pr_1)I_n(pr_1)K^\prime_n(pr_2)K_n(pr_2) \nn\\
\lo= \frac{1}{2\pi}\int_0^{2\pi}d\phi\;\frac{\pd K_0}{\pd r_1}(pR^*(r_1,r_2,\phi)) \frac{\pd K_0}{\pd
r_2}(pR^*(r_1,r_2,\phi))\;.
\end{eqnarray} 
Because the integral is convergent
we may set $\L$ to $\infty$. We find 
\begin{eqnarray} 
\fl h_2(r_1,r_2,\L,\Delta) \nn\\
\lo= \frac{\Delta^2r_1r_2}{\pi}\int_0^\infty du\;K_1(u)^2\;\int_0^{2\pi}d\phi\;
\frac{(r_1-r_2\cos(\phi))(r_2-r_1\cos(\phi))}{(r_1^2+r_2^2-2r_1r_2\cos(\phi))^{5/2}}\;,
\end{eqnarray} 
with $r_1=R-\d/2,~r_2=R+\d/2$. After manipulation we find that
\begin{equation}
h_2(r_1,r_2,\L,\Delta)~=~\frac{3}{64}\frac{\Delta^2}{R}\frac{1-y^2}{y^2}\int_0^\infty
dz\; \frac{y^2z^4-1}{(1+y^2z^2)^{1/2}(1+z^2)^{5/2}}\;, 
\end{equation} where $y =\delta/2R$. 
This gives 
\begin{equation}
h_2(r_1,r_2,\L,\Delta)~=~\frac{\Delta^2}{2\delta^2}~+~
\frac{3}{64}\frac{\Delta^2}{R}\left[\log\left(\frac{\delta}{2R}\right)
+ 2\log 2 -\frac{1}{2}\right]\;.
\label{h2final}
\end{equation}

\subsection{\label{ECE_A}$R \to \infty,$ $\d$ fixed}
To calculate $F_C(R,\delta)$ in the grand canonical ensemble we must subtract
from $F^{(0)}_{MW}(R,\delta)$ the free energy $F_\infty$, defined in
Eq. (\ref{Finf}), for a flat membrane of the same area and
thickness. This has been calculated in previous work \cite{deho} but
is it instructive to derive it directly from Eq. (\ref{Fghm}).  In the
limit $R \to \infty$ the arguments of all functions for $p \ne 0$
become large and we find that the calculation is dominated by large
$n$. The leading asymptotic results given in Eq. (\ref{ASYM}) will be
sufficient to compute $F^{(0)}_{MW}(\bs,R,\d)$ in the large $R$ limit.

From Eq. (\ref{ASYM}) we have for large $n$ that 
\begin{eqnarray}
I_n(pr)&\sim&\frac{1}{\sqrt{2\pi}}\frac{1}{(n^2+p^2r^2)^{1/4}}
\:\exp\big[n\eta(pr/n)\big]\;,\nn\\
K_n(pr)&\sim&\sqrt{\frac{\pi}{2}}\frac{1}{(n^2+p^2r^2)^{1/4}}
\:\exp\big[-n\eta(pr/n)\big]\;,\nn\\
I_n^\prime(pr)&\sim&\sqrt{\frac{1}{2\pi}}\frac{(n^2+p^2r^2)^{1/4}}{pr}
\:\exp\big[n\eta(pr/n)\big]\;,\nn\\
K_n^\prime(pr)&\sim&-\sqrt{\frac{\pi}{2}}\frac{(n^2+p^2r^2)^{1/4}}{pr}
\:\exp\big[-n\eta(pr/n)\big]\;. \label{IK}
\end{eqnarray} 
In the limit $R \to \infty$ it is better to define a new
two-dimensional wave-vector $\bk = (n/R,p)$ (i.e.,~$k_1=n/R,~k_2=p$),
since we then find that the $R \to \infty$ limit can be formulated in
terms of functions with finite arguments. The measure is then 
\begin{equation}
L\sum_n\int\frac{dp}{2\pi}~\to~2\pi LR\int
\frac{d^2k}{(2\pi)^2}\;. \label{MEASURE} 
\end{equation} 
Note that $A = 2\pi LR$ is the area of the tube.

From Eq. (\ref{IK}) and using the definition of $\eta(z)$ in
Eq. (\ref{TETA}) we obtain 
\begin{equation} 
\fl \eta\left(\frac{p}{n}(R+\d/2)\right) - \eta\left(\frac{p}{n}(R-\d/2)\right)~\sim~
k_2\d\:\eta^\prime\left(k_2/k_1\right) ~=~(k_1^2+k_2^2)^{1/2}\d~\equiv~k\d\;.
\end{equation} 
The next correction is $O(\d^3/R^2)$ which is negligible in the
$R \to \infty$ limit. Then we find 
\begin{eqnarray}
(I_n(pr_1)K_n(pr_1))^\prime
&\sim&(I_n(pr_2)K_n(pr_2))^\prime~\sim~0\;,
\label{KDEP2} \\
I_n^\prime(pr_1)I_n(pr_2)K_n(pr_1)K_n^\prime(pr_2)&\sim&-\frac{1}{4p^2r_1r_2}\exp(-2k\delta)\;. \label{KDEP4}
\end{eqnarray} 
Using Eq. (\ref{KDEP2}), we see immediately that the
contributions from the individual surfaces vanish in this limit. The
non-zero contributions then arise only from $h(r_1,r_2,\L,\Delta)$ and
trivially from $m(\L,\Delta)$ in Eq. (\ref{Ffns}). On substitution in
to Eq. (\ref{Fghm}) we find in the large $R$ limit that 
\begin{equation} 
\fl \beta F_\infty(\d)~=~-\frac{1}{2}\int
\frac{d^2k}{(2\pi)^2}\;\Bigg[\log\Big(1 - \Delta^2\Big)~-~ \log\Big(1
- \Delta^2\exp(-2k\d)\Big)\Bigg]\;. \label{F0RINFTY} 
\end{equation} 
We consider this result in the limit $\d \to 0$. The integral in
Eq. (\ref{F0RINFTY}) must be regulated with a UV cutoff $k \le
\Lambda$. The second logarithm in this equation can be expanded and
the series in $\Delta^2$ integrated term by term. If we assume that
$\Lambda\d \gg 0$ then we find 
\begin{equation} \beta
F_\infty(\d)~=~-\frac{A}{8\pi}\Bigg[\frac{\Lambda^2}{4}\log\Big(1 -
\Delta^2\Big)~-~ \frac{1}{2\d^2}\sum_m
\frac{\Delta^{2m}}{m^3}\Bigg]\;.  
\end{equation} 
The result behaves like
$1/\d^2$ but only as long as the assumption $\Lambda\d \gg 0$ holds
since terms containing the factor $\exp(-\Lambda\d)$ have been
ignored. If all terms are kept then, of course, $\lim_{\d \to
0}F_\infty(\d) = 0$.

On subtracting $2\pi RLk_BTF_\infty$ from $F^{(0)}_{MW}$ to obtain the
grand free energy $F_C$ we see that the first term in $F_\infty$
cancels the contribution from $m(\L,\Delta)$ in Eq. (\ref{Fghm})
identically.  We retain the second term in $F_\infty$ at $O(\Delta^2)$
and it cancels a similar term in the evaluation of the integral for
$h_2(r_1,r_2,\L,\Delta)$. This term is exhibited explicitly in
Eq. (\ref{h2final}).  Putting our results together we find that
$\kappa_C$ defined in Eq. (\ref{Fandkc}) is given by 
\begin{equation}
\kappa_C~=~\frac{\Delta^2}{64}\left[3\log\left(\frac{\pi\delta}{a}\right)
+ 6\log 2 + 3\gamma_E - 4\right] + \Delta^4B(\Delta)\;, \label{kappac}
\end{equation} 
where $\gamma_E$ is Euler's constant and the constant in the
brackets is evaluated to be $0.02954\ldots$.  We note an important
point which is that the $\log(R)$ dependences from the functions $g$
and $h$ cancel exactly giving a leading order behaviours of $F_C \sim
1/R$.

\section{\label{numbers}Numerical Results}
In order to calculate the Casimir energy as a function of $R$ and $\d$
we evaluate $F_C(R,\d)$, defined in Eqs.  (\ref{F0}) and (\ref{Fghm}),
numerically. The free energy $F_C(R,\d)$ is normalised to zero for
$R=\infty$, and is defined in terms of the free energy
$F^{(0)}_{MW}(R,\d)$ which is normalised to be zero when the
dielectric constant of the membrane, $\epsilon_M$ is set equal to that
for water: $\epsilon_M = \epsilon_W$. Here, $F^{(0)}_{MW}(R,\d)$ is
given as a sum over $n$ and integral over $p$ of
$F^{(0)}_{MW}(\bs,R,\d)$ ($\bs = (n,p)$), itself defined in
Eq. (\ref{F0}).

To carry out both the sum over $n$ and the integral over $p$ we use
the VEGAS integration package \cite{lepage} which is an efficient
algorithm which uses importance sampling to do multidimensional
integrals. Although we are dealing with a discrete sum over $n$ it is
easy to adapt the integrand so that it is a function of the continuous
variable $x$ through the relation 
\begin{equation} n(x)~=~R\;\mbox{Int}(\hat
x)\;,~~~~\hat x~=~x - 0.5(1-\sign(x))\;.  
\end{equation} 
Then $n(x)$ takes
integer values necessary for the summation whilst $x$ is used as a
continuous integration variable by VEGAS. Both the sum over $n$ and
the integral over $p$ are done by efficient importance sampling
techniques, and an accurate answer can be obtained. To impose the
needed Ultra-Violet regulator or cut-off, we set $\bk = (n(x)/R,\;p)$
and integrate over the region $-\pi/a \le k \le -\pi/a\;,~~k = |\bk|$.

The evaluation of the integrand poses some difficulties since, as we
have seen in the previous section, the integrand is dominated by large
values of $k$, and hence the arguments of the Bessel functions in the
definition of $F^{(0)}_{MW}(\bs,R,\d)$, Eqs. (\ref{Fghm}) and
(\ref{Ffns}), become very large indeed. In this case the function
$I_n$ ($K_n$) suffers from floating point overflow (underflow), which
can be seen easily from the asymptotic forms given in
Eq. (\ref{ASYM}).  However, in contrast the products over the Bessel
functions which constitute each term in Eq. (\ref{Ffns}) do not suffer
in this way. This also can be seen from Eq. (\ref{KDEP4}) where the
increasing and decreasing exponential behaviours of $I_n$ and $K_n$,
respectively, compensate to give the behaviours $\exp(-2k\d)$. To
construct a robust integrand we used routines for the full Bessel
functions when $k$ was sufficiently small and used the appropriate
asymptotic form given in (\ref{ASYM}) when either $n$ or $p$, or both,
became large. It was then possible to cancel the diverging and
vanishing exponential factors against each other, so obtaining a well
defined integrand computationally.

We take $\epsilon_W/\epsilon_M = 40$ and evaluate $F_C(R,\d)$ as a
function of $R$ in nanometers, and $\d = 5(nm),~10(nm)$ and for
various values of the cutoff length $a$. Because there is no
electrolyte the temperature dependence is purely in the factor of
$k_BT$ multiplying our calculation. From Eq. (\ref{kappac}) it is
clear that, as is true in most applications, the Casimir energy is
dominated by the UV cutoff behaviour and hence by the value chosen for
$a$. It is not fully clear what the correct value for $a$ should be
since the microscopic properties of the membrane interface are not
properly included in the analysis. Typically, we would expect $a$ to
be the scale of the inter-molecular spacing of the molecules forming
the membrane or of water molecules. For this calculation a reasonable
value is $a \sim 0.5(nm)$. To test the validity of our UV analysis we
first investigate how $F_C(R,\d)$ behaves for very small $a$ and we
choose $a=0.05,~0.1,~0.2,~0.5(nm)$.  The function $B(\Delta^2)$
receives contribution from the both the $g$ and $h$ terms in
Eq. (\ref{F}) with $B(0) \ne 0$. Note that there are no odd terms in
$\Delta$ in the leading $1/R$ behaviour of $F_C$ since to leading order
one may set $\delta/R=0$ (equivalently $R_1=R_2$) in the leading order
behaviour of $g$ and in the denominator of the second term in the
logarithm of the integral defining $h$.  This is a consistent
parametrisation whilst $\delta \gg a$. The limit $\delta \to 0$ must
be taken carefully and when $\delta < a$ the separation of
$F^{(0)}_{MW}$ in Eq. (\ref{Fghm}) into contributions from functions
$g$ and $h$ is not useful since $h$ develops the compensating UV
divergence to that in $g$ and we find $\lim_{\delta \to 0}F_C = 0$, as
expected; in essence, the larger of $(\delta,a)$ acts as the $UV$
cut-off on the integral defining $h$.

In Table \ref{table}, for various values of $\Delta$ and $\delta/a$,
we compare the prediction of Eq. (\ref{kappac}) with the result of
numerical evaluation  and deduce a numerical value for $B(\Delta^2)$. Owing to
small systematic errors in the numerical calculation of the Bessel
functions there is a tiny discrepancy for very small $\Delta$ but
$B(\Delta^2)$ is seen to be a constant function from evaluations at
larger $\Delta$ and we see that $B(0)$ is plausibly $1/8$.

\btab 
\begin{center}
\btabu{|c|c|c|c|c|}\hline
$\Delta$&$~~\delta/a$~~&\parbox{30mm}{$O(\Delta^2)$ coeff. of $1/R$\\
from Eq. (\ref{kappac})}&\parbox{25mm}{Coeff. of $1/R$\\from
simulation}&$B(\Delta^2)$\\\hline
78/82&$10^3$&-0.342&-0.443&0.123\\\hline
78/82&$10^2$&-0.244&-0.346&0.123\\\hline
0.6&$10^3$&-0.1361&-0.1520&0.123\\\hline
0.6&$10^2$&-0.0972&-0.0162&0.123\\\hline
0.2&$10^3$&-0.0151&-0.0162&--\\\hline
0.6&$10^3$&-0.0038&-0.0040&--\\\hline \etabu
\caption{\label{table}
For various values of $\Delta$ and $\delta/a$
we compare the prediction of Eq. (\ref{kappac}) with the result of
simulation and deduce a numerical value for $B(\Delta^2)$. Owing to
small systematic errors in the numerical calculation of the Bessel
functions there is a negligible discrepancy for very small $\Delta$
but $B(\Delta^2)$ is seen to be a constant function from evaluations
at larger $\Delta$. We see that the result for $F_C$ from
Eq. (\ref{kappac}) is in very good agreement with the full
calculation. Various values of $\delta$ and $a$ were used but
typically $\delta = 1-10$(nm)}\end{center}  
\etab

The physical value of the UV cut-off length can only be determined
phenomenologically in this model. This is because the model is an
effective field theory in which the dynamics of the molecular electric
dipoles is described by the dielectric constant which is a static
long-range parameter. The field modes with large-$k$ and $n$ probe the
static short distance properties of the model and so a more refined
field theory is needed for these scales. It is unclear whether the
molecular nature of the lipid has an effect on the UV cutoff but it
would seem most likely that the effective value of $\epsilon_W$ at
short scales are dominant in this calculation. The effect is encoded
in the value of $a$.

\section{\label{conclusion} Conclusion}

In this paper we have developed the theory for a general approach to
the calculation of the electrostatic free energy for a system of
symmetrically layered electrolytic membranes. The definition of
symmetrical is that the Laplacian is separable and that the coordinate
direction normal to the layers can be interpreted in terms of a
Euclidean time variable, $t$.  The x partition function for a layer
bounded by surfaces defined by $t=t_i$ and $t=t_j,~t_i < t_j$ can then
be written in terms of the Feynman (Euclidean) time evolution kernel
from $t_i$ to $t_j$ by invoking the well-known connection between this
formalism and statistical mechanics. The method is a general extension
of our work concerning flat membranes \cite{deho},\cite{dehown},
\cite{dehocon}, \cite{dehotloop} and allows the full interacting
Sine-Gordon field theory in Eq. (\ref{S_L}) for electrolytic layers to
be studied systematically, including the perturbation theory in the
coupling constant $g$ defined by $g=l_B/l_D$, where $l_B$ and $l_D$
are the Bjerrum and Debye lengths, respectively.  Geometries of
interest to which this approach applies include cylindrical and
spherical ones.  In this paper the general theory developed in section
\ref{theory} is applied to the system of cylindrical layers each
filled with a pure dielectric medium where the dielectric constant
differs between layers. The analysis of this system is based on the
free harmonic field theory which is exactly soluble and which is the
theory about which a perturbative expansion for the effect of
fluctuations takes place. In the succeeding sections the particular
problem of the free energy of a tube of dielectric material submersed
in water is studied and the Casimir force calculated. The tube is a
simple model for the t-tubule formed from lipid membrane in particular
kinds of muscle cell and the object of the calculation is to
investigate the size and form of the Casimir force and whether it can
act to stabilise the tube against bending stress that acts to decrease
the tube curvature and hence increase its radius. In order to
calculate the relevant free energy we assumed that the tube was
connected to an infinite reservoir of flat membrane and so worked with
the grand ensemble describing this system. The assumption about which
ensemble is the relevant one is crucial to answering questions
concerning tubule stability since in the grand ensemble the nature of
the lipid or membrane reservoir determines its bulk free energy and
hence affects the energy of conformation of the tube; a flat membrane
reservoir will differ for a reservoir which stores spare lipid at the
end of the tubule essentially as crumpled membrane. Clearly also, if
there is no reservoir at all, so that the ensemble is canonical, then
any increase in the surface area of the tube as it expands will lead
to a lower density of lipids in the surface and a concomitant increase
in the surface energy over and above that due to the bending stress;
this is a fundamentally different situation to the one we assume.

In the case studied here we see from Table \ref{table} that for a
lipid bilayer tube in water with $\delta=10$(nm) and $a=0.1$(nm) we
find $\kappa_C = 0.346$. It is possible to include the contributions
from the modes with non-zero Matsubara frequencies, a calculation in
progress, but we can expected at most a factor of two or so
enhancement on past experience of similar calculations (\cite{mani}),
and so $\kappa_C \sim 1$ is a likely largest value . These are at the
lowest end of values for $\kappa_B$ for known lipid bilayers in water
(\cite{boal,wurg}). However, W\"urger (\cite{wurg}) calculates
$\kappa_B$ for surfactant films, analysing the role of hydrophobic
tails, as a function of the tail length and the area per molecule, and
finds a wide range of values for $\kappa_B$ ($\kappa_B = \pi \kappa$,
with $\kappa$ from ref. (\cite{wurg})) including values small enough,
corresponding to soft interfaces, to accommodate our result. Thus it
is conceivable that there can be small tubes formed from soft
membranes in water for which the bending forces tending to expand the
radius are compensated by the Casimir attraction, and the tube is
stabilised by sub-leading $O(1/R^2)$ forces.

\appendix
\section{Asymptotic behaviour of modified Bessel functions}
For completeness we include the asymptotic behaviours of 
the modified Bessel functions $I_n$ and $K_n$ as given in (\cite{abst}).
For $z\to 0$:
\begin{eqnarray}
I_n(z)&\sim& \frac{(z/2)^n}{n!} \nn\\
K_n(z)&\sim&\frac{1}{2}n!(z/2)^{-n} \ (n>0)\nn\\
K_0(z)&\sim& -\{\log(z/2)+\gamma\}I_0(z).
\end{eqnarray}

For $z\to \infty$:
\begin{eqnarray}
 I_n(z)&\sim&\sqrt{\frac{1}{2\pi
z}}e^z\left\{1-\frac{\mu-1}{8z}+\frac{(\mu-1)(\mu-9)}{2!(8z)^2}-
\ldots\right\}\nn \\
K_n(z)&\sim&\ds \sqrt{\frac{\pi}{2z}}e^{-z}\left\{1+\frac{\mu-1}{8z}+
\frac{(\mu-1)(\mu-9)}{2!(8z)^2}-\ldots\right\} ,
\end{eqnarray}
where $\mu = 4n^2$.

For $n\to \infty$:
\begin{eqnarray}
I_n(nz)&\sim& \frac{1}{\sqrt{2\pi
n}}\frac{e^{n\eta}}{(1+z^2)^{1/4}}
\left\{1+\sum_{k=1}^\infty\frac{u_k(t)}{n^k}\right\} \nn \\
K_n(nz)&\sim&
\frac{\pi}{\sqrt{2n}}\frac{e^{-n\eta}}{(1+z^2)^{1/4}}
\left\{1+\sum_{k=1}^\infty(-1)^k\frac{u_k(t)}{n^k}\right\}
\nn \\  
I_n^\prime(nz)&\sim& \frac{1}{\sqrt{2\pi
n}}\frac{(1+z^2)^{1/4}}{z}e^{n\eta}
\left\{1+\sum_{k=1}^\infty\frac{v_k(t)}{n^k}\right\} \nn
\\  K_n^\prime(nz)&\sim&
-\frac{\pi}{\sqrt{2n}}\frac{(1+z^2)^{1/4}}{z}e^{-n\eta}
\left\{1+\sum_{k=1}^\infty(-1)^k\frac{v_k(t)}{n^k}\right\},
\label{ASYM} 
\end{eqnarray} 
where
\begin{equation}
t~=~\frac{1}{\sqrt{1+z^2}}\;,~~~~~~\eta~=~\sqrt{1+z^2} +
\log\frac{z}{1+\sqrt{1+z^2}}\;, \label{TETA} 
\end{equation} 
and $u_k(t),~v_k(t)$
are polynomials of order $3k$ and are even (odd) if $k$ is even (odd)
\cite{abst}.

\section*{References}
\bibliography{refs} 
\bibliographystyle{unsrt}

\end{document}